# Optical Wireless Cabin Communication System

Osama Zwaid Alsulami
*School of Electronic and Electrical Engineering*
*University of Leeds*
Leeds, United Kingdom
ml15ozma@leeds.ac.uk

Mohammed T. Alresheedi
*Department of Electrical Engineering*
*King Saud University*
Riyadh, Saudi Arabia
malresheedi@ksu.edu.sa

Jaafar M. H. Elmirghani
*School of Electronic and Electrical Engineering*
*University of Leeds*
Leeds, United Kingdom
j.m.h.elmirghani@leeds.ac.uk

*Abstract*— Offering a high data rate to users inside the cabin is attractive for the aircraft building companies. This paper presents an optical wireless system that uses visible light for transmitting data. The reading light for each seat is used as a transmitter in this work to offer high data rates. Red, yellow, green, and blue (RYGB) laser diode (LD) are utilised in each reading light to obtain a high modulation bandwidth. Two types of reading lights based on angle diversity transmitter (ADT) units are utilised in this paper for illumination and communication. Additionally, two kinds of optical receivers are used: four branches angle diversity receiver (ADR) and 25 pixels imaging receiver (ImR). The delay spread and the signal to interference and noise ratio (SINR) are evaluated. The proposed system can offer high data rates up to 22.8 Gbps for each user by using simple on-off-keying (OOK) modulation.

## I. INTRODUCTION

We are currently witnessing a growing demand for bandwidth and high data rates inside the aircraft by using wireless communication. As a result, the future technology of wireless communication must support these demands. Radio frequency (RF) is the current wireless technology used . It has some limitations such as the capacity of the channel and the achievable transmission rate, limited by the availability of the radio spectrum. Also, it may interfere with another system that uses the same spectrum in the aircraft. In addition, extremely high data rates for multi-users that are above 10 Gbps and into the Tbps regime are difficult to achieve using the bandwidth available to radio systems. Cisco expects that the number of users using the Internet will increase 27 times between 2016 to 2021 [1]. Optical wireless communication (OWC) is a promising technique that can increases the achievable data rates and service quality. Visible light communication (VLC) is one of the possible realisations of Optical wireless (OW) communication. It can offer a license free bandwidth, low cost components and high security when compared to RF systems [2]-[9]. However, some limitations exist when using VLC systems, and one of the main limitations is the potential absence of line-of-sight (LOS) components. This weakens the system's performance considerably [10], [11]. In addition, interference between users and the inter-symbol interference (ISI) from multi-path propagation may reduce the system's performance. Many studies [12]-[21] on VLC systems show that, video, data and voice can be transferred through VLC systems at data rates up to 20 Gbps in indoor environments. Thus, VLC systems can be used on board aircrafts to connect passengers to the Internet. VLC systems use light for data transfer, thus, VLC systems can offer a solution for the aircraft instruments' incompatibility with RF systems and the potential interference caused by RF systems.

In this paper, a VLC system for use inside the aircraft is introduced. A light engine that uses red, yellow, green, and blue (RYGB) Laser Diodes (LDs) is used as a VLC transmitter. In addition, two kinds of optical receivers are considered in this work: an angle diversity receiver (ADR) and an imaging receiver (ImR). The effects of multi-path propagation are considered in the developed model. By using RYGB LDs, two advantages can be achieved (1) a white colour can be introduced to provide illumination for the indoor environment as stated in [22], also, (2) a high modulation bandwidth for data transmission can be provided. This paper is organised as follows: The system configuration is described in Section II and the optical receiver design is shown in Section III. Section IV introduces the optical transmitter design. In Section V, the simulation results are shown, and the conclusions are provided in Section VI.

## II. SYSTEM CONFIGURATION

The aircraft considered in this work is boing 747-400. In addition, the evaluation is carried out in the main deck that consists of 539 economy class seats. The aircraft cabin dimensions (length × width × height) are 57 m × 6.37 m × 2.41 m as shown in Fig. 1 based on [23]. In the simulation, the aircraft internal surface was divided into small areas to simplify the computation. In this work, first and second order reflections were considered, noting that reflections higher than second order have little influence on the received optical power [24], [25]. MATLAB was used in the simulation in conjunction with the ray-tracing algorithm for calculating the reflections from ceiling, walls and the floor inside the aircraft. Therefore, each surface inside the aircraft was divided into small equal areas, $dA$, with a reflection coefficient of $\rho$. It is shown in [25] that plaster walls and walls with similar surface roughness reflect light rays close to a Lambertian pattern. Thus, each element ($dA$) in each surface (ceiling, walls and floor) inside the aircraft is modelled as a Lambertian reflector with a reflection coefficient equal to 0.8 for ceiling and aircraft walls (walls next to seats 1 and 10), and 0.3 for the floor and other surfaces. Elements in each surface were considered as small emitters that reflect the received signal in the shape of a Lambertian pattern with emission order equals to 1. The element's area has an inverse relationship with the temporal resolution of the impulse response results. Therefore, by decreasing the area of the element, the resolution of the results improves. However, reducing the area of the elements increases the computation time of the simulation. Therefore, the area of the element for the first order reflection was chosen to be 5 cm × 5 cm, while the element's area for the second order reflection was selected to be 20 cm × 20 cm, and that can keep the computation time of the simulation within a reasonable time [14], [26]. The areas under the seats surface are blocked and this means all communication is done above the seats surface.

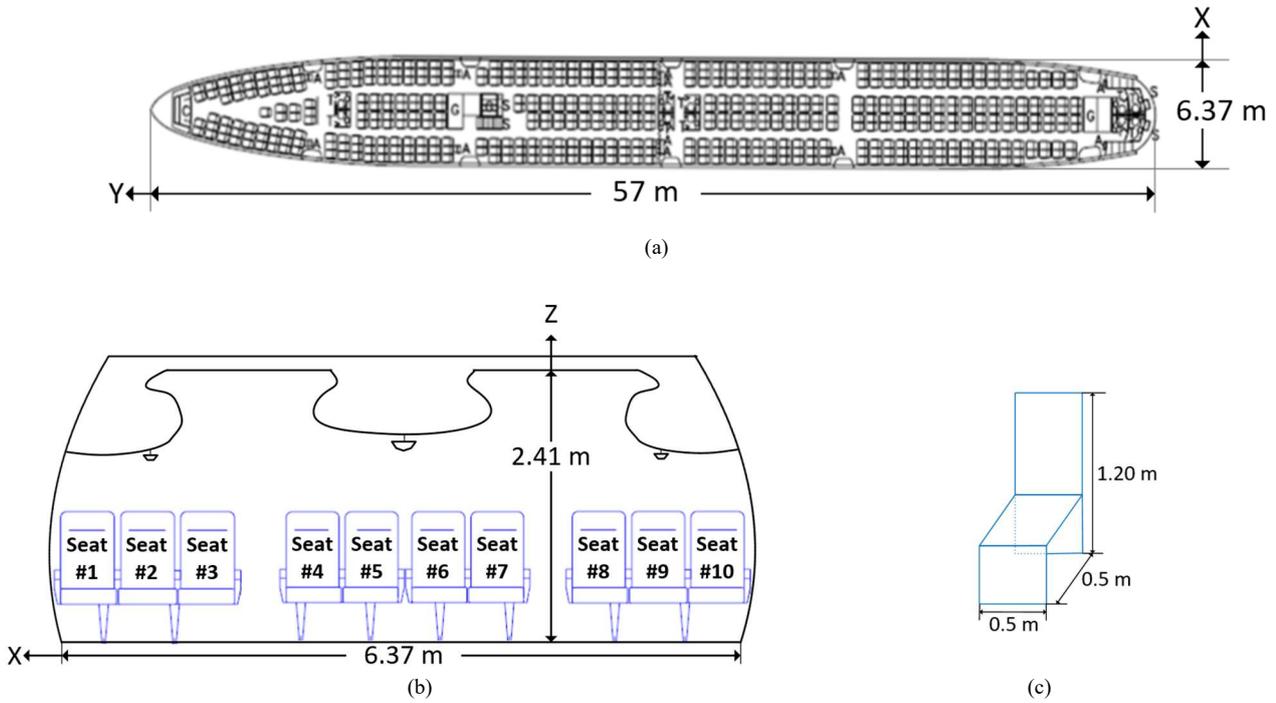

Figure 1. System Configuration: (a) aircraft dimension, (b) aircraft dimension, (c) seat dimension.

### III. OPTICAL RECEIVER DESIGN

In this work, two kinds of optical receiver designs have been examined: an angle diversity receiver (ADR) and an imaging receiver (ImR). The ADR has 4 branches of detectors with narrow field of views (FOVs) (See Fig. 2a). The narrow FOVs collect the signals while reducing the delay spread and potentially co-user interference. Each detector is oriented at a specific area that differs from the others and thus covers different areas in the ceiling based on two angles: elevation ($El$) and azimuth ($Az$) angles. The $El$ angles of these detectors set at 70°. While, the $Az$ angles of these detectors are chosen to be 45°, 135°, 225° and 315°. The FOV of these detectors is selected to be 21°. Additionally, each detector has an area equal to 4 mm$^2$ with responsivity equal to 0.4 A/W. The second optical receiver is the ImR. The ImR contains 25 pixels and each one of these pixels has an area equal to 4 mm$^2$ and responsivity equal to 0.4 A/W with a narrow FOV. A lens was used above all the pixels. The imaging receiver has a FOV equal to 40° and collects light rays from a large area see Fig. 2b.

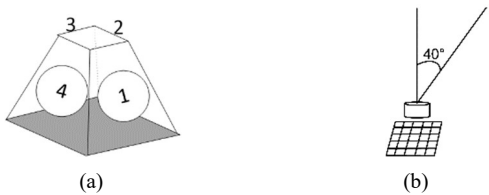

Figure 2. Optical Receiver Design: (a) ADR, (b) ImR.

### IV. OPTICAL TRANSMITTER DESIGN

In this paper, the reading light for each seat is used as a visible light communication (VLC) transmitter. Two types of reading light units were used and each unit uses an angle diversity transmitter (ADT). The first reading light unit is used above the seats that placed next to the aircraft wall. It consists of 3 branches and each branch containing an RYGB LD light unit. Each branch has a narrow beam that covers a different seat inside the aircraft. The beam's size can be adjusted and is specified by the beam Lambertian emission order ($n$). If $n$ is large, the size of the beam is small which also reduces the coverage area of the branch and that can help reduce the interference and beam intersections for beams generated by the different branches. As a result, the semi angle of each branch is chosen to be 14º and hence each branch covers only one seat. The reading light unit is located on the middle of the ceiling of the three seats next to the wall (see Fig. 1b, seats numbers:1,2,3,8,9,10). Each branch is oriented to a different seat based on two angles $El$ and $Az$. The $El$ angles of the three branches are set as follow: two branches are set at -55°, while the branch that faces downwards to cover the middle seat is oriented at -90°, whereas, the branches' $Az$ angles are set equal to 180°, 0° and 0° respectively. The second reading light unit, which is located above the 4 middle seats (see Fig. 1b, seats numbers: 4,5,6,7) inside the cabin, contains 4 branches with each branch having RYGB LD. Each branch has a narrow beam that covers a different seat (the beam semi angle is thus selected). For this light reading unit, the semi angles of the branches are chosen to be 10° which results in each branch covering only one seat. The reading light unit is positioned on the middle of the ceiling of these seats. Each branch covers a different seat based on $El$ and $Az$ angles. The $El$ angles of these branches are set at -65°, -80°, -80° and -65 and the $Az$ angles of these branches are set equal to 180°, 180°, 0° and 0° respectively.

### V. SIMULATION SETUP AND RESULTS

The bit error rate (BER) is a function of the signal to interference and noise ratio (SINR) and for on-off keying (OOK) systems, it can be calculated as:

$$BER = Q(\sqrt{SINR}), \quad (1)$$

where Q(·) is the Q-function and is computed as follows:

$$Q(x) = \frac{1}{2} erfc\left(x/\sqrt{2}\right) \approx \frac{1}{\sqrt{2\pi}} \frac{e^{-(x^2/\sqrt{2})}}{x} \quad (2)$$

The SINR at the seats for both types of optical receiver ADR and ImR is expressed as:

$$SINR = \frac{R^2 (P_{s1}-P_{s0})^2}{\sigma_t^2 + \sum_{i=1}^{I} R^2 (P_{i1}-P_{i0})^2} \quad (3)$$

where i is the number of reading lights sources that interfere with the current reading light source, R is the detector responsivity which is equal to 0.4 A/W in this work for both types of receivers, $P_{s1}$ is the received power related to logic 1, $P_{s0}$ is the received power related to logic 0, $P_{i1}$ is the interfering power received from the other reading light sources related to logic 1, $P_{i0}$ is the interfering power received from the other reading lights sources related to logic 0 and $\sigma_t$ is the total noise of the received signal and is expressed as follows:

$$\sigma_t = \sqrt{\sigma_{pr}^2 + \sigma_{bn}^2 + \sigma_{sig}^2} \quad (4)$$

where, $\sigma_{pr}$ is the noise of the preamplifier, $\sigma_{bn}$ is the ambient noise and $\sigma_{sig}$ is the noise associated with the received signal. Two methods are used to combine the received signals from detectors of the ADR or pixels of the ImR. These are selection combining (SC) and Maximum ratio combining (MRC). Both methods are evaluated and compared for both types of receivers. The SINR for the SC method is computed as:

$$SINR_{SC} = max_k \left(\frac{R^2 (P_{s1}-P_{s0})^2}{\sigma_t^2 + \sum_{i=1}^{I} R^2 (P_{i1}-P_{i0})^2}\right)_k, \quad 1 \leq k \leq J \quad (5)$$

where $J$ is the total number of photodetectors or pixels, while, the SINR for the MRC method is calculated as:

$$SINR_{MRC} = \sum_{k=1}^{J} \left(\frac{R^2 (P_{s1}-P_{s0})^2}{\sigma_t^2 + \sum_{i=1}^{I} R^2 (P_{i1}-P_{i0})^2}\right)_k, \quad 1 \leq k \leq J \quad (6)$$

The ImR that is used in this work has one lens above all the pixels with a transmission factor based on the incidence angle ($\Upsilon$). The transmission factor of the lens used is given by [27]:

$$Tc(\Upsilon) = -0.1982\Upsilon^2 + 0.0425\Upsilon + 0.8778 \quad (7)$$

Fig. 3 illustrates the delay spread and the SINR of the two receiver types. Fig. 3a compares the delay spread of the ImR and the ADR. The ImR provides a lower delay spread compared to the ADR. However, the delay spread of both types of receivers can support a high channel bandwidth which can result in a higher data rate. This work shows that the available channel bandwidth can support data rates up to 22.8 Gbps using OOK modulation. The SINR for both kinds of receivers using the two combining methods is shown in Fig. 3b. The ImR can provide a stable SINR for all seat locations compared to the ADR which results in a higher SINR in some user locations. This is due to the narrower FOVs of the ImR compared to the ADR which reduces the interference between all seat locations. In the ImR, the SC and the MRC methods show the same result since one pixel can see the transmitter, while in the ADR, more than one detector can see the transmitter. Thus, the MRC method when using ADR offers a better SINR than the SC method. The reason behind that, the MRC method is able to maximise the SINR by adding all outputs that come from all detectors together.

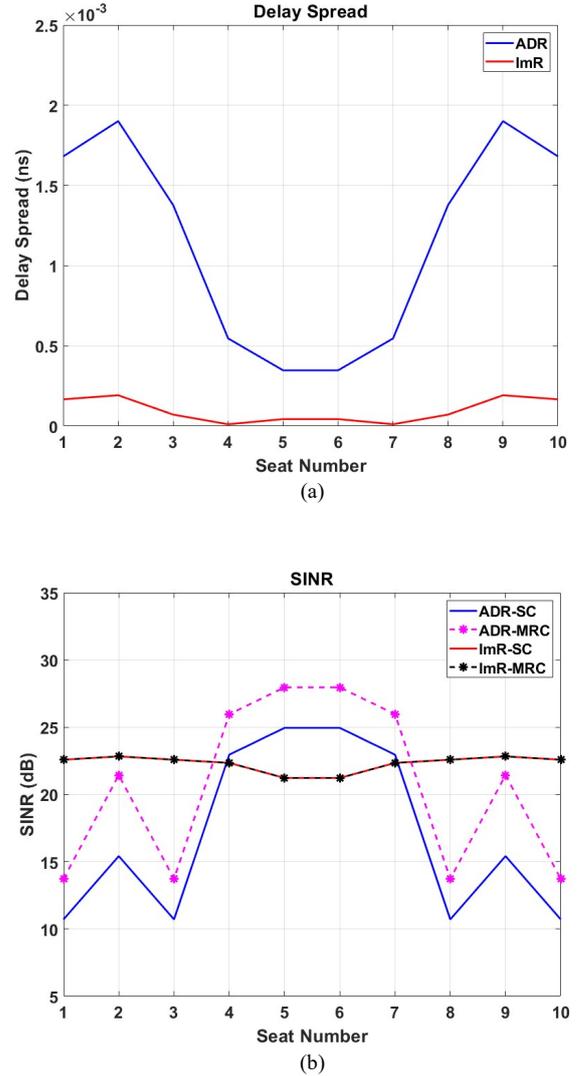

Figure 3. (a) Delay Spread, (b) SINR.

## VI. CONCLUSIONS

This paper proposed an on-board OWC system for boing 747-400 using two types of reading lights unit that utilise angle diversity transmitters. The two types of reading light units were used for reading illumination and communication (infrared can be used for communication if light dimming is needed or when the lights are switched off). In addition, two kinds of optical receivers were used in this work: four branch ADR and 25 pixels imaging receiver. The delay spread and the SINR were determined. The two types of receivers were compared in this paper. The proposed systems can provide a high SINR and data rate for different seat locations inside the cabin using simple OOK modulation. Both types of receivers offer a high SINR supporting a high data rate for each user up to 22.8 Gbps.

ACKNOWLEDGMENT

The authors would like to acknowledge funding from the Engineering and Physical Sciences Research Council

(EPSRC) for the TOWS project (EP/S016570/1). The first author would like to thank Umm Al-Qura University in the Kingdom of Saudi Arabia for funding his PhD scholarship. All data are provided in full in the results section of this paper.